\documentclass[review]{elsarticle}

\pdfoutput=1

\usepackage[utf8]{inputenc}
\usepackage{rotating}

\usepackage{booktabs}
\usepackage{wrapfig}
\usepackage{graphicx}
\usepackage{pdfpages}
\usepackage{url}
\usepackage{breakurl}
\usepackage[breaklinks]{hyperref}

\usepackage{lineno,hyperref}
\usepackage{amsmath}

\journal{Journal of Epidemiology, Elsevier}

\begin{document}

\begin{frontmatter}

\title{Modelling the impact of vaccination on the COVID-19 pandemic in African countries}

\author[add1]{Dephney Mathebula\corref{cor1}}
\ead{mathed2@unisa.ac.za}
\author[add2]{Abigail Amankwah}
\author[add3]{Kossi Amouzouvi}
\author[add4]{K\'et\'evi A. Assamagan\corref{cor1}}
\ead{ketevi@bnl.gov}
\author[add5]{Somi\'ealo  Azote}
\author[add6]{Jesutofunmi Ayo Fajemisin}
\author[add7]{Jean Baptiste Fankam Fankam}
\author[add8]{Aluwani Guga}
\author[add9]{Moses Kamwela}
\author[add10]{Toivo S. Mabote}
\author[add11]{Mulape M Kanduza}
\author[add10]{Francisco Fenias Macucule}
\author[add8]{Azwinndini Muronga}
\author[add12]{Ann Njeri}
\author[add13]{Michael Oluwole}
\author[add10]{Cl\'audio Mois\'es Paulo}
\cortext[cor1]{Corresponding Authors}
\address[add1]{University of South Africa, Department of Decision Sciences, South Africa}
\address[add2]{African Institute for Mathematical Sciences(AIMS)--Ghana }
\address[add3]{Kwame Nkrumah University of Science and Technology, Ghana  \& Technical University of Dresden, Germany}
\address[add4]{Brookhaven National Laboratory, Physics Department, Upton, New York, USA}
\address[add5]{Universit\'e de Lom\'e, D\'epartement de Physique, Lom\'e, Togo}
\address[add6]{University of South Florida, Department of Applied Physics Tampa, Florida, USA}
\address[add7]{University of Yaounde I,  Department of Physics,Yaounde, Cameroon}
\address[add8]{Nelson Mandela University, South Africa}
\address[add9]{Lusaka Apex Medical University, Zambia}
\address[add10]{Universidade Eduardo Mondlane, Grupo de Astrofísica, Ci\^{e}ncias Espaciais e Intelig\^{e}ncia Artificial, Maputo, Mozambique}
\address[add11]{Cancer Diseases Hospital, Lusaka, Zambia}
\address[add12]{University of Manchester, UK}
\address[add13]{University of Ibadan, Nigeria}

\begin{abstract}

The rapid development of vaccines to combat the spread of COVID-19 disease caused by the SARS-CoV-2 virus is a great scientific achievement. Before the development of the COVID-19 vaccines, most studies capitalized on the available data that did not include pharmaceutical measures. Such studies focused on the impact of non-pharmaceutical measures (e.g social distancing, sanitation, wearing of face masks, and lockdown) to study the spread of COVID-19. In this study, we used the SIDARTHE-V model which is an extension of the SIDARTHE model wherein we include vaccination roll outs. We studied the impact of vaccination on the severity (deadly nature) of the virus in African countries. Model parameters were extracted by fitting simultaneously the COVID-19 cumulative data of deaths, recoveries, active cases, and full vaccinations reported by the governments of Ghana, Kenya, Mozambique, Nigeria, South Africa, Togo, and Zambia. With countries having some degree of variation in their vaccination programs, we considered the impact of vaccination campaigns on the death rates in these countries. The study showed that the cumulative death rates declined drastically with the increased extent of vaccination in each country; while infection rates were sometimes increasing with the arrival of new waves, the death rates did not increase as we saw before vaccination. 
\end{abstract}


\end{frontmatter}


\newpage
\section{Introduction}
\label{sec:intro}

\noindent Since 2019, the severe acute respiratory syndrome coronavirus~2 (SARS-CoV-2) that causes the coronavirus disease 2019 (COVID-19) has been spreading worldwide \cite{lai2020severe}.  To mitigate the spread of COVID-19, different control measures such as lockdown, social distancing, wearing of face masks, sanitising, vaccination were implemented and that led to the reduction in the number of new COVID-19 cases. Studies on the impact of vaccination, more especially in South Africa, Brazil and Germany, showed that there is a direct relation between the number of vaccinated individuals and the number of new COVID-19 cases, i.e. the higher the number of vaccinated individuals, the lower the number of new COVID-19 cases~\cite{ribeiro2022characterisation, AMOUZOUVI2021e00987, mukandavire2020quantifying, kassa2021modelling}. Nevertheless, some studies confirmed that despite the effectiveness of COVID-19 vaccination and treatment in reducing the spread, non-pharmaceutical control measures should continue to be implemented~\cite{diagne2021mathematical}. 

In this study, we investigated the impact of vaccination during the second year since the emergence of COVID-19 in seven African countries, namely Ghana, Kenya, Mozambique, Nigeria, South Africa, Togo, and Zambia. This study is a continuation of our work reported in Ref.~\cite{AMOUZOUVI2021e00987}. COVID-19 transmission rate across Africa varied over time because of changes in behaviours and government policies as the pandemic evolves, and the introduction of vaccination programs. As a result, we modelled the outbreak in the aforementioned African countries, considering that several model parameters varied over time.

We analyzed data taken over a two-year period: the first (without vaccination) and second (with vaccination) years of the pandemic. We extracted and compared parameters to gauge the impact of vaccination as a pharmaceutical intervention.  A few COVID-19 vaccines have been developed by pharmaceutical and biotech companies. The vaccines differ in their development, efficacy, storage and administration. The commonly used vaccines are from Pfizer and Johnson \& Johnson. Our study of vaccination impact was informed by the vaccination programs in the African countries considered. 

In developed countries, general vaccination roll outs started much earlier in 2021 while most of the African countries started vaccination campaigns later. Because of the unequal availability of vaccines around the world, the starting dates of the vaccination might have an impact on this study. However, given that we have analyzed vaccination data for one year in all countries considered, the starting dates of vaccination do not affect our conclusions. 

\noindent The structure of this paper is as follows. We present the formulation of the SIDARTHE-V model in Section~\ref{sec:sidarthev}, taking the influence of vaccination campaigns into consideration in Ref.~\cite{Giordano2021}. In Section ~\ref{sec:analysis}, we discuss the analysis of COVID-19 data with vaccination campaigns in the African countries taken into consideration in this study. In Section ~\ref{sec:disc}, we discuss the impact of vaccination, and in Section ~\ref{sec:conc}, we provide our concluding remarks.

\section{SIDARTHE-V model with vaccination roll outs}
\label{sec:sidarthev}

\noindent In this study, we applied the SIDARTHE-V model~\cite{Giordano2021} with vaccination campaigns in the second year of the pandemic. The original SIDARTHE-V of ~\cite{Giordano2021} assumes that all vaccinated are immunized.  In this study, we considered the possibility that vaccinated individuals can still get infected, and become infectious; these dynamics are captured by connecting the $V$ and $I$ compartments, as shown in Figure~\ref{fig:SIDARTHEV}, where the parameters and variables of the model are presented. Equations~\ref{EQ01} describe the pandemic evolution, with vaccination roll outs:
\begin{eqnarray}
 \left\{\begin{array}{lcl}
\dot{S} &=& - \left(\alpha I + \beta D + \gamma A + \delta R\right)S - \phi S\\
\dot{V}&=& - \alpha' IV + \phi S \\
\dot{I} &=& \left(\alpha I + \beta D + \gamma A + \delta R\right)S + \alpha'IV - \left(\epsilon + \lambda + \zeta\right)I\\
\dot{D} &=& \epsilon I - \left(\eta + \rho\right) D\\
\dot{A} &=& \zeta I - \left(\theta + \mu + \kappa\right)A\\
\dot{R} &=& \eta D + \theta A - \left(\tau_1 + \nu \right)R\\
\dot{T} &=&  \mu A + \nu R - \left(\tau_2 + \sigma\right) T \\
\dot{H} &=& \lambda I + \kappa A + \sigma T + \xi R + \rho D\\
\dot{E} &=& \tau_1 R + \tau_2 T
\label{EQ01} 
 \end{array}\right.
 \end{eqnarray}

The basic reproduction number, $R_0$, is the average number of secondary cases produced by an infected individual in a population where everyone is susceptible~\cite{van2002reproduction}. Estimating $R_0$ helps in the implementation of appropriate responses to pandemic evolution, in particular, the number of people to vaccinate for herd immunity. For the SIDARTHE-V model, Equations~\eqref{EQ01}, the $R_0$ is given by: 
\begin{eqnarray}
R_0 &=& \displaystyle{\frac{\alpha r_{2} r_{3} r_{4} + \beta \epsilon r_{3} r_{4} + \delta \epsilon \eta r_{3} + \delta r_{2} \tau \zeta + \gamma r_{2} r_{4} \zeta}{r_{1} r_{2} r_{3} r_{4}}},
\label{EQ02} 
  \end{eqnarray}
 where $r_1 = \epsilon + \zeta + \lambda, ~r_2 = \eta + \rho, ~ r_3 = \theta + \mu + \kappa, ~r_4 = \nu + \xi$. For a better understanding of the $R_0$ derivation, see~\cite{Giordano2020}. From Equation~\eqref{EQ02}, it can be seen that $R_0$ depends on the model parameters that affect pandemic evolution. Thus, it is very important to understand the model parameters and to make sure they are extracted correctly.

\section{Analysis of COVID-19 data with vaccination}
\label{sec:analysis}

\noindent In our previous work~\cite{AMOUZOUVI2021e00987}, we studied the impact of non-pharmaceutical control measures on the spread of COVID-19 in African countries was considered,  Nigerian COVID-19 data was not included. For this reason, we start this section with the analysis of the data of Nigeria from the time when the first COVID-19 case was identified in that country. This includes the first year with no vaccination followed by another year with vaccination roll outs. For all the African countries considered in this study, our model starts from the first reported COVID-19 case in each country throughout the first year without vaccination into the second year of data with vaccination. 

\subsection{Analysis of COVID-19 data of Nigeria}

\noindent In Nigeria, they confirmed the first case in the Infectious Disease Centre, Yaba, Lagos State, on February 27, 2020. An airplane from Milan, Italy,  arrived at the International Airport, Lagos, on February 14, 2020, with an infected Italian citizen who went to his company's site in Ogun State the following day. The health authorities (Nigeria Centre for Disease Control) implemented containment measures by contact tracing of ‘Persons of Interest’ which included all persons on the flight and those he had close contact with while in Lagos and Ogun States~\cite{NCBI2020}. After a period of two weeks, cases were detected in Lagos and Abuja and this marked the emergence of the spread in the country. The Federal Government restricted international commercial flights into the country, effective from March 23, 2020~\cite{NCAA2020}.

The Federal Government ordered the closure of schools and all non-essential services (businesses and industries) and ordered the cessation of all movements in Lagos State, Ogun State, and the Federal Capital Territory, Abuja, on March 29, 2020, for an initial period of 14 days. Later, the restriction on movements was extended for another 14 days from April 12, 2020~\cite{NCDC2020}.

Most State Governments restricted public gatherings and religious activities for over fifty persons. The Federal Government lifted the travel ban on domestic flights on April 20, 2020, and ordered a nationwide overnight curfew on movements from 8:00~pm to 6:00~am on May 2, 2020, and later eased the overnight curfew on movements on September 3, 2020, to be from 12:00~am to 4:00~am.

On May 4, 2020, the Federal Government authorized the gradual easing of lockdown in the previously restricted states and mandated the use of face masks in public. 

On May 6, 2020, the Federal Government announced an extension of the travel ban on both international and local flights to, June 7, 2020, to curb the spread of coronavirus in the country.

The Federal Government reopened international flights for operations on August 29, 2020~\cite{NCDC2022}. On January 27, 2021, the President signed six COVID-19 Health Protection Regulations 2021, with restrictions on gatherings, operations of public places, mandatory compliance with treatment protocols, offenses and penalties, enforcement and application, and lastly the interpretation and citations of the regulations~\cite{NCDC2021}.

After the first confirmed case on February 27, 2020, the number of confirmed cases increased drastically and the total number of confirmed cases as of March 27, 2022, was 255,341 with a total number of 249,566 discharged cases and 2,633 active cases. The first death case was on March 23, 2020; death cases have increased to a total number of 3,142 as of March 27, 2022. The health sector started the COVID-19 sample test on April 8, 2020, and on March 27, 2022, they recorded total tests of 4,589,725~\cite{NCAA2020,NCDC2020}.

The first shipment of four million Oxford-AstraZeneca COVID-19 vaccines arrived in the country on March 2, 2021, and vaccination began on March 5, 2021. The country received subsequent shipments of Moderna, Johnson \& Johnson, and Pfizer COVID-19 vaccines on August 1, August 12, and October 14, 2021, respectively. Due to the single dose requirement of Johnson \& Johnson COVID-19 vaccine, Nigeria's National Primary Health Care Development Agency (NPHCDA) prioritised hard-to-reach and vulnerable areas for vaccination~\cite{VON2021}. 

As of March 27, 2022, there were 21,049,754 persons who have received their first dose and 9,565,143 who have received their second dose~\cite{VON2021}.

\begin{figure}[!htbp]
 \begin{center}
  \includegraphics[width=\textwidth]{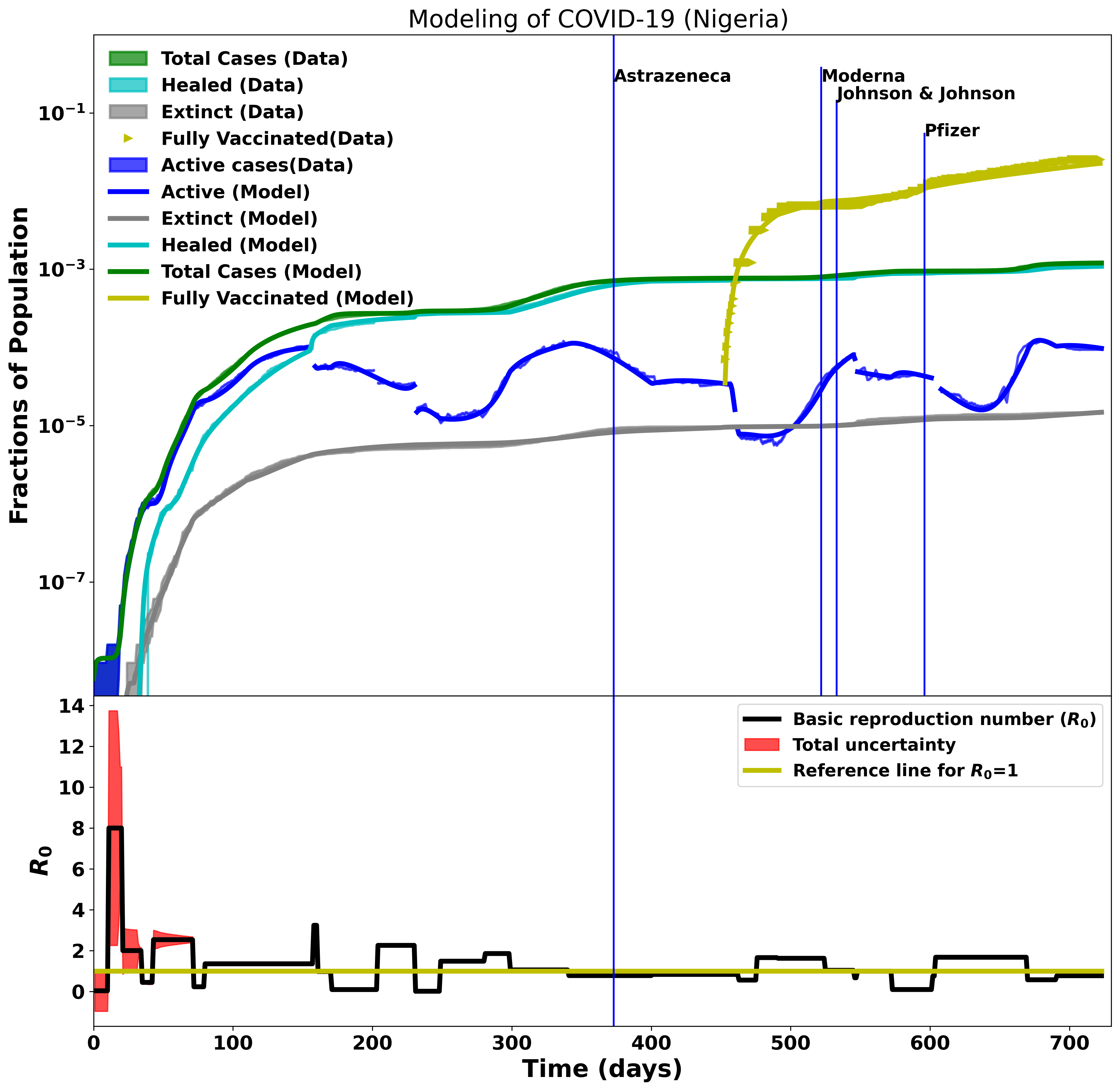}
   \caption{The modelling of 2 years of COVID-19 data of Nigeria. Day~0 corresponds to the onset of the pandemic, i.e.  February 27, 2020. The top plot shows the data and model for active, recovered, death, total cases, and fully-vaccinated individuals. The Vaccination drive started on March 5, 2021. The bottom plot shows the time-dependent basic reproduction number.}
    \label{fig:Nigeria1}
  \end{center}
\end{figure}

Figure~\ref{fig:Nigeria1} (top plot) shows the SIDARTHE-V modelling of the Nigerian COVID-19 data of active, recovered, extinct and fully-vaccinated cases. The time-dependent basic reproduction number $R_0$, obtained by fitting the model to the data, is shown in the bottom plot.

The $R_0$ increased significantly to eight after a week. This was largely due to the learning period about the pandemic and the lack of public control measures.

Around day 35, the $R_0$ dropped below one mainly because of the introduction of public control measures by the government and awareness by the public. Another increase in $R_0$ to a point above two was observed around day 40 most likely because of the difficulties to comply with the control measures.

Around day 65, it also dropped below one. The $R_0$ later increased around day 75 above one and later rose to a point above three around day 150 due to the ineffectiveness of the measures in some parts of the country and the lack of enforcement strategies from the government.

Around day 165, the $R_0$ dropped well below one and increased above two around day 205. Another drop occurred around day 230 to point zero after some restrictions from the government. We see that around day 250, there was an increase in $R_0$ above one and it was within the range of two around day 280. Even after day 700, $R_0$ remained below two. These fluctuations were due to the negligence of the people to observe the control measures.

Figure~\ref{fig:Nigeria2} shows the quality of the modelling as ratios of data over model predictions; the figure also shows the model prediction of the infected but unaffected population.

The vaccination has eased the anxiety caused by the pandemic and also enabled the government to relax lockdown protocols. Businesses and institutions such as the education sector have resumed their services.

\subsection{COVID-19 vaccination analysis for South Africa}
\label{sec:sa}

\noindent In South Africa, COVID-19 vaccination has been an ongoing immunisation campaign to vaccinate 40 million South Africans~\cite{Vaccine2022}. Four types of COVID-19 vaccines were approved by the South African Health Products Regulatory Authority (SAHPRA), namely, Johnson \& Johnson, Pfizer, Sinovac, and AstraZeneca~\cite{Vaccine2022}. For the South Africa COVID-19 case study, Johnson \& Johnson's Janssen and Pfizer vaccines were considered~\cite{SA2022}. As of June 9, 2022, $535,714$  COVID-19 hospital admissions were recorded in South Africa~\cite{HOSPADM2022}. 

In our previous study~\cite{AMOUZOUVI2021e00987}, we covered the South African COVID-19 data up to adjusted alert level~3 that was an effect from December 29, 2020, to February 28, 2021~\cite{AMOUZOUVI2021e00987}. Based on the changes of COVID-19 new cases in South Africa, the government introduced adjusted alert levels, defined in Ref.~\cite{AMOUZOUVI2021e00987}, as follows~\cite{SAlifted2022, Vaccine2022}:
\begin{itemize}
\item Level 1: March 1--May 30, 2021;  
\item Level 2: May 31--June 15, 2021;
\item Level 3: June 16--June 27, 2021;
\item Level 4: June 28--July 25, 2021;
\item Level 3: July 26--September 12, 2021;
\item Level 2: September 13--30, 2021; and 
\item Level 1: October 1, 2021--April 14, 2022.
\end{itemize}
On May 3, 2022, South Africa confirmed $3,661,635$ recovered individuals, $100,377$ death cases and $\sim17.7$ million vaccinated individuals, and $3,802,198$ positive cases~\cite{Vaccine2022}. The National State of Disaster in South Africa has been lifted since April 5, 2022~\cite{SAlifted2022}.

In South Africa, the healthcare workers were the first group to be vaccinated; it started on February 18, 2021 (day 350) until May 17, 2021 (day 439) under phase~1 of the Sisonke Protocol, which enabled the government to make the Johnson \& Johnson vaccine quickly accessible through a research initiative~\cite{PETER20222, Sisonke}. The death case remained constant during phase 1 while the number of active, healed, and total cases slightly remained constant.  During Phase 2 which started on May 18, 2021, everyone from age 16 and above was allowed to be vaccinated with the first dose of Johnson \& Johnson and Pfizer.

\begin{figure}[!h]
 \begin{center}
 	\includegraphics[width=\textwidth]{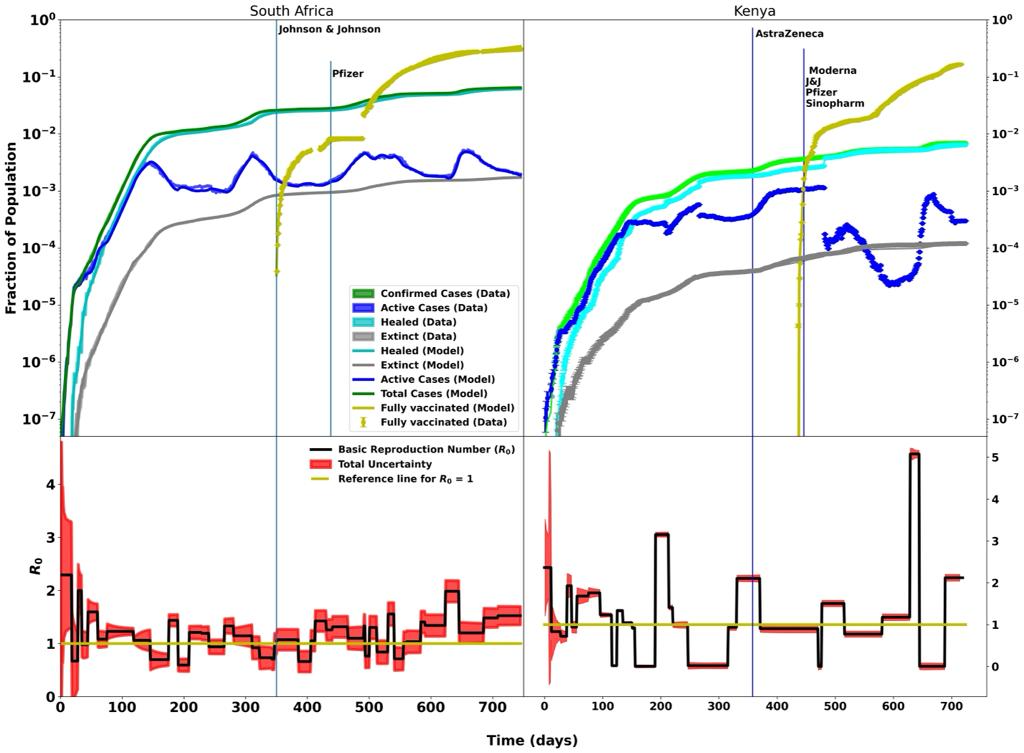}
  	\caption{The modelling of about 2 years of COVID-19 data of South Africa (left plot) and Kenya (right plot). Day~0 corresponds to the onset of the pandemic, i.e.  March 5, and March 12, 2020, for South Africa and Kenya respectively. The top plots show the data and model for active, recovered, death, and total cases, and for fully-vaccinated individuals. Vaccination drives started on February 28, 2021 (South Africa) and March 5, 2021 (Kenya). The bottom plots show the time-dependent basic reproduction numbers.}
   	\label{fig:SouthAfricaKenya}
  \end{center}
\end{figure}

Figure~\ref{fig:SouthAfricaKenya} (left plots) shows the modelling of the South African data; the first year of the pandemic was studied and discussed in Ref.~\cite{AMOUZOUVI2021e00987}. The second year of the South African COVID-19 data, with vaccination roll outs, is extensively discussed in Section~\ref{sec:disc}.

\subsection{COVID-19 vaccination analysis for Kenya}

\noindent The data used in this analysis were taken from the daily press releases by the Ministry of Health, Government of the Republic of Kenya~\cite{GoK}. Having received the first 1.12 million doses of the Oxford-AstraZeneca vaccine, the vaccination drive in Kenya kicked off on March 5, 2021. This was one year after the first COVID-19 case was reported on March 12, 2020. Six hundred and sixty-seven doses of AstraZeneca were administered on the first day of vaccination to front-line healthcare workers only at the Kenyatta National Hospital, Nairobi. This was then followed by other essential workers such as security officers and teachers in the first few weeks of the vaccination program, followed by targeted people with higher risks of severe disease and those aged 50 years and above. The second dose was administered on May 28, 2021, and 203 people received their second dose. 

After five months of administering the AstraZeneca vaccine, 880,460 doses of the Moderna vaccine were received on August 23, 2021, from the US government via COVAX, making Moderna the second COVID-19 vaccine to be offered in the country. Additional 141,600 doses of Johnson \& Johnson vaccine were received soon afterward on September 3, 2021. This was the third vaccine type to be offered and totaled to 4.2 million doses of vaccine received~\cite{GoK}. On September 17, 2021, the country received 795,600 doses of the Pfizer vaccine from the US government, making Pfizer the fourth vaccine offered. Shortly afterward, on September 18, 2021, 200,000 doses of Sinopharm COVID-19 vaccine were received from the Chinese government. The government had authorised all five vaccines and at the time of writing, they were being used across the country.

After a slow uptake of the vaccines among the population due to vaccine hesitancy~\cite{Orangi2021}, a spike was observed on November 23, 2021, with the highest number of vaccination doses administered to 103,506 people in a single day. This followed a government directive on November 21, 2021, stating that anyone not vaccinated by December 21, 2021, would be refused in-person government services and access to public entertainment spots such as restaurants. As of December 2021, 7\% of the population was fully vaccinated and $\sim 10\%$ of the population was partly vaccinated. This figure slightly surpassed the government target of 10 million people by the end of the year.  Figure~\ref{fig:SouthAfricaKenya} (right plots) shows the modeling of two years of COVID-19 data in Kenya with the vaccination rolls commencing on day 358 (highlighted by the blue vertical line), almost a year after the first COVID-19 case was reported---a detailed study of the data before vaccination campaigns were discussed in Ref.~\cite{AMOUZOUVI2021e00987}. The issuance of the second dose began around day 450 as highlighted by the second blue vertical line. Around day 480 ($\sim 30$ days after the second dose), there was a sharp decrease in active cases. Into the second year of the COVID-19 pandemic, the basic reproduction number $R_0$ remained $\approx$1 or below 1 with slight variations during minor peaks. At day $\sim$ 650, $R_0$ increased sharply to $\sim 5$. This was due to a slight but sharp increase in active cases, following a steady decrease in active cases in the country. 

Kenya is part of the WHO AFRO 20 priority African countries with the slowest rates of COVID-19 vaccination uptake~\cite{WHO_ke}. Therefore, the WHO AFRO implemented phased COVID-19 vaccination campaigns in February 2022 to boost vaccination rates. This entailed community outreach efforts and increased vaccination sites from 800 to 6,000 sites. Over two weeks (February 3--17), the daily vaccination average increased from 70,000 to 200,000 people. This also raised the percentage of the population that was fully vaccinated from 9.9\% to 13.4\%. As of March 11, 2022, two years after the first COVID-19 case was reported and one year after the mass vaccination program roll out, 8,054,405 vaccine doses were administered and $\sim 14.8\%$ (7,930,000) of the population was fully vaccinated. At the time of writing, a total of 323,140 COVID-19 cases had been reported and a total of 5,644 deaths recorded.

COVID-19 restrictions are no longer in place though the government is encouraging citizens to wear masks and maintain social distancing where possible. Factors affecting the vaccination program in Kenya include: i) funding, ii) the availability of vaccines, iii) storage requirements, iv) vaccine hesitancy among the population~\cite{Orangi2021} and geographical inequalities in accessing vaccines in hard-to-reach areas~\cite{MUCHIRI2022}. The government aims to vaccinate 15.91~million people by June 2023 in a 3-phased roll-out approach initially targeting 1.25~million people by June 2021 in phase one. This was followed by phase two, July 2021--June 2022, with a target of 9.76~million people, including the elderly and people with underlying health conditions. The third phase started in July 2022 and will run until June 2023, with a target of 4.9~million people above 18 years old, those with underlying health risks, and essential workers.

\subsection{COVID-19 vaccination analysis for Ghana}

\begin{figure}[!h]
 \begin{center}
 	\includegraphics[width=\textwidth]{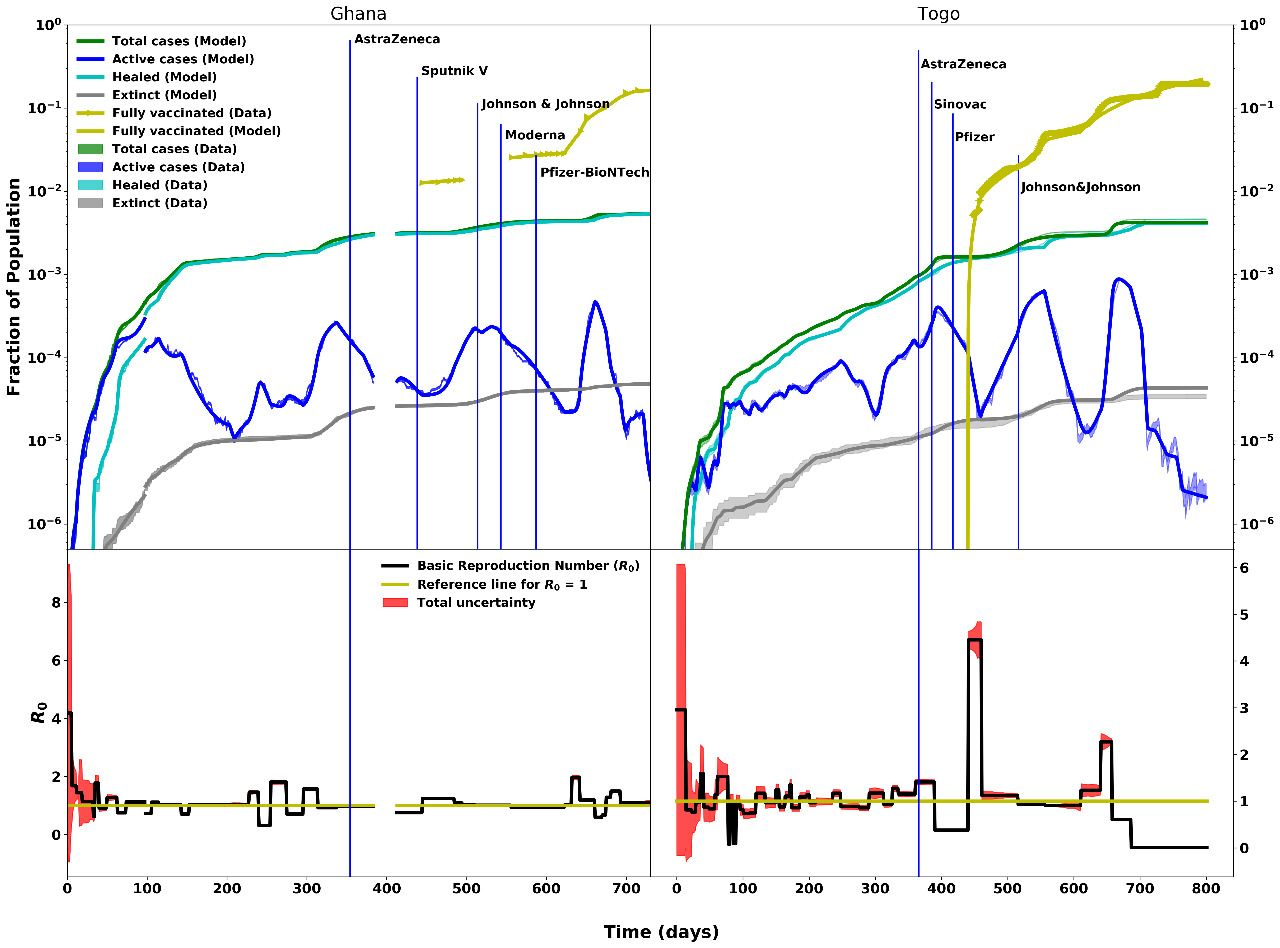}
  	\caption{The modelling of about 2 years of COVID-19 data of Ghana (left plot) and Togo (right plot). Day~0 corresponds to the onset of the pandemic, i.e.  March 12, and March 6, 2020, for Ghana and Togo respectively. The top plot shows the data and model for active, recovered, death, and total cases, and for fully-vaccinated individuals. Vaccination drives started on March 1, 2021 (Ghana) and March 9, 2021 (Togo). The bottom plots show the time-dependent basic reproduction numbers.}
   	\label{fig:GhanaTogo}
  \end{center}
\end{figure}

\noindent In Ghana, the government committed to acquiring COVID-19 vaccines on December 20, 2020~\cite{lamptey2021nationwide}. The arrival of the first shipment of COVID-19 vaccines, from the COVAX initiative to African countries, enabled Ghana Health Authority to soon begin its first vaccine roll out on March 1, 2021~\cite{WHOGH, Ghana-COVAX, nonvignon2022estimating} with the AstraZeneca vaccine.  Johnson \& Johnson (J\&J), Moderna, Pfizer, and Sputnik~V were the COVID-19 vaccines also approved and administered in Ghana. Figure~\ref{fig:GhanaTogo} (left plots) shows the modelling of the Ghanaian data over a two-year period: data from the first year of the pandemic---before vaccination started---were analyzed and discussed in Ref.~\cite{AMOUZOUVI2021e00987}; in this study, we focused on the second year of data with vaccination drives.

The four reported COVID-19 waves in Ghana were caused by the emergence of Beta, Alpha, Delta and Omicron coronavirus variants. These waves are characterised by the four peacks in the top-left plot of Figure~\ref{fig:GhanaTogo}. A study conducted in Ref.~\cite{morang2022genetic} indicated that Delta and Alpha were among the most viral variants in Ghana. At the time of writing, the Beta variant was still being monitored in Ghana since it had the third-highest frequency. During the second wave, regions further from Accra, such as the Northern and Upper East, had different variants. These locations lagged behind the rest of the country in the third wave and did not appear to experience one~\cite{G.H.S}. The Beta variant was prominent in Ghana when the airport reopened to foreign travelers in September 2020, and its related reproduction number started decreasing from August 2020 as shown in the bottom-left plot of Figure~\ref{fig:GhanaTogo}. The Alpha variant superseded Beta in January 2021 and became the major cause of all reported illnesses until June 2021, when Delta lineages took over. The Delta lineages started in June 2021 until December 2021. The major coronavirus variants were first detected in traveller samples, before in community instances~\cite{morang2022genetic}.

The president of Ghana and his vice were the first to receive the AstraZeneca vaccine on March~1, 2021~\cite{CNR}. By March~2,  2021, vaccination was launched in the Ashanti region and over 10,000 people had been vaccinated. The second dose of the AstraZeneca vaccine commenced on May~19, 2021.

By April~25, 2022, $14,268,269$ doses of these vaccines have been administered; $18.3\%$ of Ghana's population has been fully vaccinated, $29.9\%$ has received at least one dose of the vaccines, and $360,201$ people have received the first booster dose. By April 30, 2022, there were $161,216$ cases in Ghana. Out of these, there were $159,737$ recoveries, $1,445$ deaths, and $34$ active cases. Great Accra region alone reported 56.3\% of the total national cases. It is followed by the Ashanti region with 13.8\% of the total cases~\cite{G.H.S}. 

\subsection{COVID-19 vaccination analysis for Togo}

\noindent On March 7, 2021, approximately one year after the detection of the first case, the country received 156000 doses of AstraZeneca through the COVAX facility~\cite{world2021covid, Konu2021.04.20.21254863}, and the vaccination campaign started the following day. 120000 additional doses of AstraZeneca were received on March 31, 2021. After these, additional 100620 Pfizer doses were obtained in May 2021, followed by 200000 doses of Sinovac on April 23, 2021. On August 7, 2021, Togo received additional 118000 doses of Johnson \& Johnson vaccine out of the 4 million doses that it had ordered. The  World Health Organisation Coronavirus  Dashboard indicates that, by August 14, 2022, Togo had received 3262548 COVID-19 vaccine doses, with 2152846 people vaccinated---corresponding to $\sim 25.4$\% of the population qualified for vaccination---and 1425113  persons fully vaccinated~\cite{WHO-Dashboard}. The vaccination started with health workers on March 10, 2021, day 370 as shown in  Figure~\ref{fig:GhanaTogo} (right plots), followed by clinically vulnerable individuals, then people over 50 years old~\cite{world2021covid, Konu2021.04.20.21254863}. It took approximately 2 months to cover this targeted population. After priority groups had been vaccinated, there was a wider roll out among younger age groups. One month after the vaccination campaign (from day 400) began, we started to see the impact on infection rate, and this is reflected in $R_0$ as shown in Figure~\ref{fig:GhanaTogo} (right plot). The data from the first year of the pandemic---before vaccination started---were analyzed and discussed in Ref.~\cite{AMOUZOUVI2021e00987}.

Active cases continued to decrease up to three months after the vaccination started while $R_0$ sharply increased in the third month. This increase in $R_0$ resulted from the relaxation of the control measures that were in place before the start of the vaccination. These measures were largely no longer respected, as people thought that the problem of COVID-19 would be solved immediately by the arrival of the vaccines. After day 470, the active cases started to increase again when the vaccine doses were finished and a new COVID-19 variant (Delta) emerged. As the active cases started to increase, the government warned the population of the new variant and encouraged rigorous adherence to the control measures. More vaccines were received later and distributed across the country. However, as the government accelerated the vaccination campaign, vaccine hesitancy set in. There was an increase in general vaccine hesitancy but especially towards COVID-19 vaccines~\cite{gittings2021even,  alemayehu2022determinants, adunimay2022western}. Measures to encourage vaccination were therefore put in place, such as obligatory presentation of the COVID-19 vaccination card before entering any public institution. Despite these different strategies, as of September 17, 2021, the proportion of the population who had received two doses of the COVID-19 vaccine was only 5.6\%. To reach the vaccination targets, the WHO Country Office in Togo provided technical and financial support to the Togolese government; through the Ministry of Health, Public Hygiene and Universal Access to Health Care (MSHPAUS), they initiated community dialogues and broad awareness-raising in the Grand-Lomé region, the epicenter of the epidemic in Togo. This reduced misinformation and removed barriers to vaccine acceptance. However, there have been rises and falls in the basic reproduction number as shown in Figure~\ref{fig:GhanaTogo} (bottom-right plot); the rises may be related to the non-respect of the control measures. This overall observation allows us to stress that both control measures and vaccination are necessary to overcome the COVID-19 pandemic.

\subsection{COVID-19 vaccination analysis for Mozambique}

\noindent The datasets used in this study for the particular case of Mozambique were taken from the daily press releases and daily bulletins on the website of the government~\cite{Moz1, Moz2}.

\begin{figure}[!h]
 \begin{center}
 	\includegraphics[width=\textwidth]{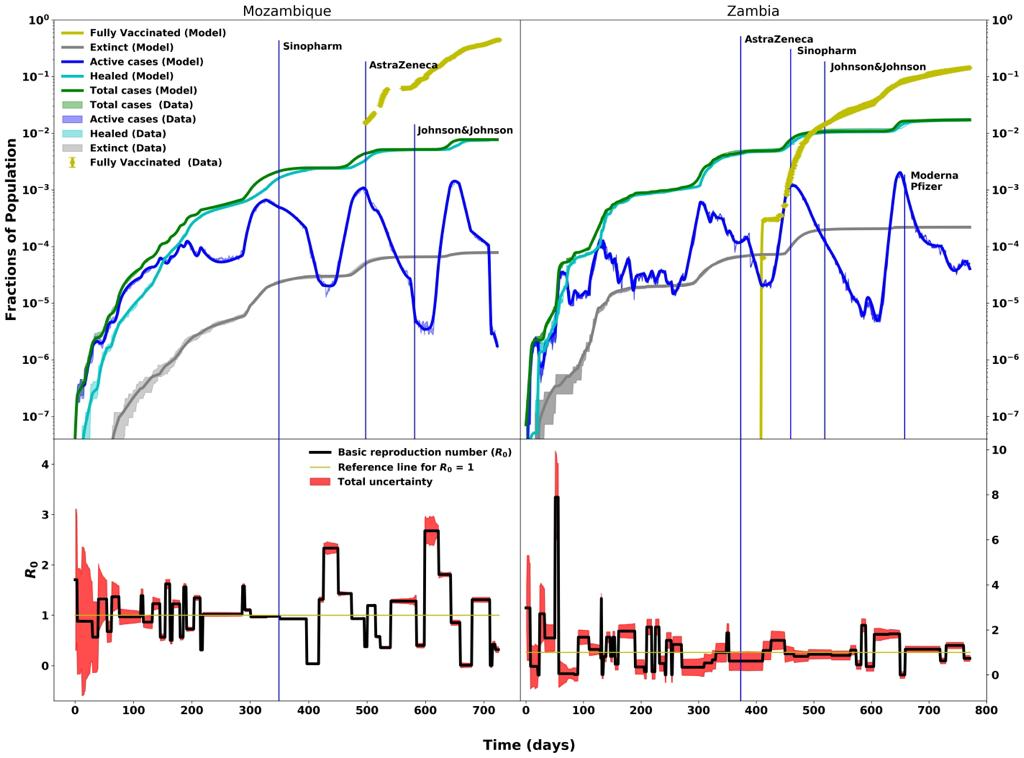}
  	\caption{Plots of $\sim$2 years modeling of COVID-19 data taking into account the vaccination campaign in Mozambique (left) and Zambia (right). Day~0 corresponds to the onset of the pandemic, i.e.  March 20, and March 18, 2020, for Mozambique and Zambia respectively. Following similar procedures from the previous works done by many of the current study's authors, see e.g. Ref.~\cite{AMOUZOUVI2021e00987}, the top plots present data and models for some of the main pandemic parameters analyzed (e.g active, recovered, death and total cases) including fully-vaccinated individuals. Vaccination drives started on March 8, 2021 (Mozambique) and April 14, 2021 (Zambia). The bottom plots show the time-dependent basic reproduction numbers.}
   	\label{fig:MozambiqueZambia}
  \end{center}
\end{figure}

Modelling of COVID-19 data was carried out in this work with the main purpose of understanding the vaccination impact during the pandemic evolution in the country. Figure~\ref{fig:MozambiqueZambia} (left plots) show the obtained results for approximately one year of the vaccination campaign. The data of the first year of the pandemic---before vaccination started---were analyzed and discussed in Ref.~\cite{AMOUZOUVI2021e00987}; Figure~\ref{fig:MozambiqueZambia} (top-left plots) also show the first year of data before vaccination.

In Mozambique, the vaccination campaign started on March 8, 2021. During this period, there was already a reduction of active cases because of non-pharmaceutical measures which were being implemented according to the Decree 7/2021 of March~5 (see Ref.~\cite{Moz3}). In general, the first vaccination campaign targeted health professionals, older people, diabetic patients, defence and security forces as well as university teachers~\cite{Moz4}. Between April 19 and May 10, 2021, Mozambique had the second stage of vaccination that covered final-year medical students, teachers who were not covered in the first stage, inmates, police, and primary school teachers. The third stage of vaccination was between October 20 and November 3, 2021; it covered carriers, people that were not vaccinated in the first two stages, motorcycle taxis, students, and all vulnerable people. Around the end of the fourth wave, on January 23, 2022, booster doses were introduced~\cite{Moz5}.

Figure~\ref{fig:MozambiqueZambia} (bottom-left plot) shows the $R_0$ evolution in Mozambique. The $R_0$ vary from 2.5 to 0.1 as follows: 1) in the second wave, the $R_0$ varied between 0.1 to 2.1; 2) in the third wave, the $R_0$ varied between 0.4 to 2.5; 3) in the fourth wave, the $R_0$ varied between 0.1 to 1.8. The $R_0$ fluctuations were related to the Government regulations of non-pharmaceutical interventions together with the onset of new variants which triggered new waves. 

During the vaccination campaign, infection was still spreading, but with diminishing impact as shown in Figure~\ref{fig:MozambiqueZambia}---the fifth wave of COVID-19 started in the last week of May 2022 and was fading at the time of writing. The onset of this wave, relatively small (duration and impact) compared to the previous ones, coincided with the time when the winter brought unusually low temperatures in some regions and many people suffered from typical flu symptoms. The rate of deaths in this wave was very low; the rate of recovery was high with a small number of people needing hospitalization.

It is important to note that the Mozambican government had set a vaccination target of 15 million people at the time of writing. In this period, about 97$\%$ were fully vaccinated---of these, 614842 individuals received the booster. The government was planning to start vaccinating people aged between 12 and 17 years~\cite{Moz1,Moz2}.

\subsection{COVID-19 vaccination analysis for Zambia}

\noindent The Zambian data of the first year of the pandemic---before vaccination started---were analyzed and discussed in Ref.~\cite{AMOUZOUVI2021e00987}. The Government of Zambia, through the Ministry of Health (MoH), officially launched the COVID-19 vaccination campaign on April 14, 2021, at the University Teaching Hospital (UTH) in Lusaka \cite{Zambia-MoH}. Zambia received more than 10 million doses of vaccines (Pfizer, Moderna, Johnson \& Johnson, Sinopharm, AstraZeneca, and others) through the COVID-19 Vaccine Global Access (COVAX) program~\cite{Zambia-MoH, Zambia-Can, Zambia-Fra}. The vaccines were distributed to various vaccination centers across the country through efficient logistics and supply chain management systems.   

The campaign for the administration of second doses of AstraZeneca and Sinopharm, and Johnson \& Johnson was initiated by July 2021. By April 30, 2022, more than a million doses of COVID-19 vaccines were administered to eligible persons (above 18 years) and priority was given to high-risk groups such as the elderly (above 65 years) and people with underlying disease(s) \cite{ Zambia-sit}. However, Zambia, like other developing countries, experienced significant vaccine hesitancy. To overcome this challenge and achieve 70\% herd immunity, the government, through MoH, introduced a door-to-door vaccination campaign and community sensitization on the benefits of vaccination~\cite{Zambia-sit}. In addition, non-pharmacological interventions such as keeping a 1-meter social distance, wearing face masks, and hand sanitizing were mandatory in public places. By the time of writing, the 70\% vaccination target was not achieved.

Figure~\ref{fig:MozambiqueZambia} (right plots) shows the progression of COVID-19 during the two years of the pandemic in Zambia. The period before the vaccination rollout was discussed in Ref.~\cite{AMOUZOUVI2021e00987}. The modelling of COVID-19 vaccination in Zambia to eligible persons began after April 14, 2021. By May 25, 2021, a total of 5286 were fully vaccinated with Sinopharm and AstraZeneca. In addition, by January 2, 2022, eligible individuals started to receive the booster vaccines, corresponding to a cumulative total of 1649. However, in about 8 to 9 months following the start of vaccination, as shown in Figure~\ref{fig:MozambiqueZambia} (bottom-right plot), $R_0$ increased twice to approximately 2 due to vaccine hesitancy, lack of strict adherence to COVID-19 protocols such as wearing of masks, hand sanitizing and keeping a safe social distance of about one meter. Finally, around day 700, $R_0$ reduced significantly to approximately one following a reduction in active cases and increased recovery. This significant reduction could be attributed to effective measures introduced by the government to curb the spread of COVID-19 such as the mandatory wearing of masks in public places, maintenance of safe distance and lockdown measures, and closure of schools and universities.

\section{Impact of vaccination}
\label{sec:disc}
\begin{figure}[!h]
 \begin{center}
 	\includegraphics[width=\textwidth]{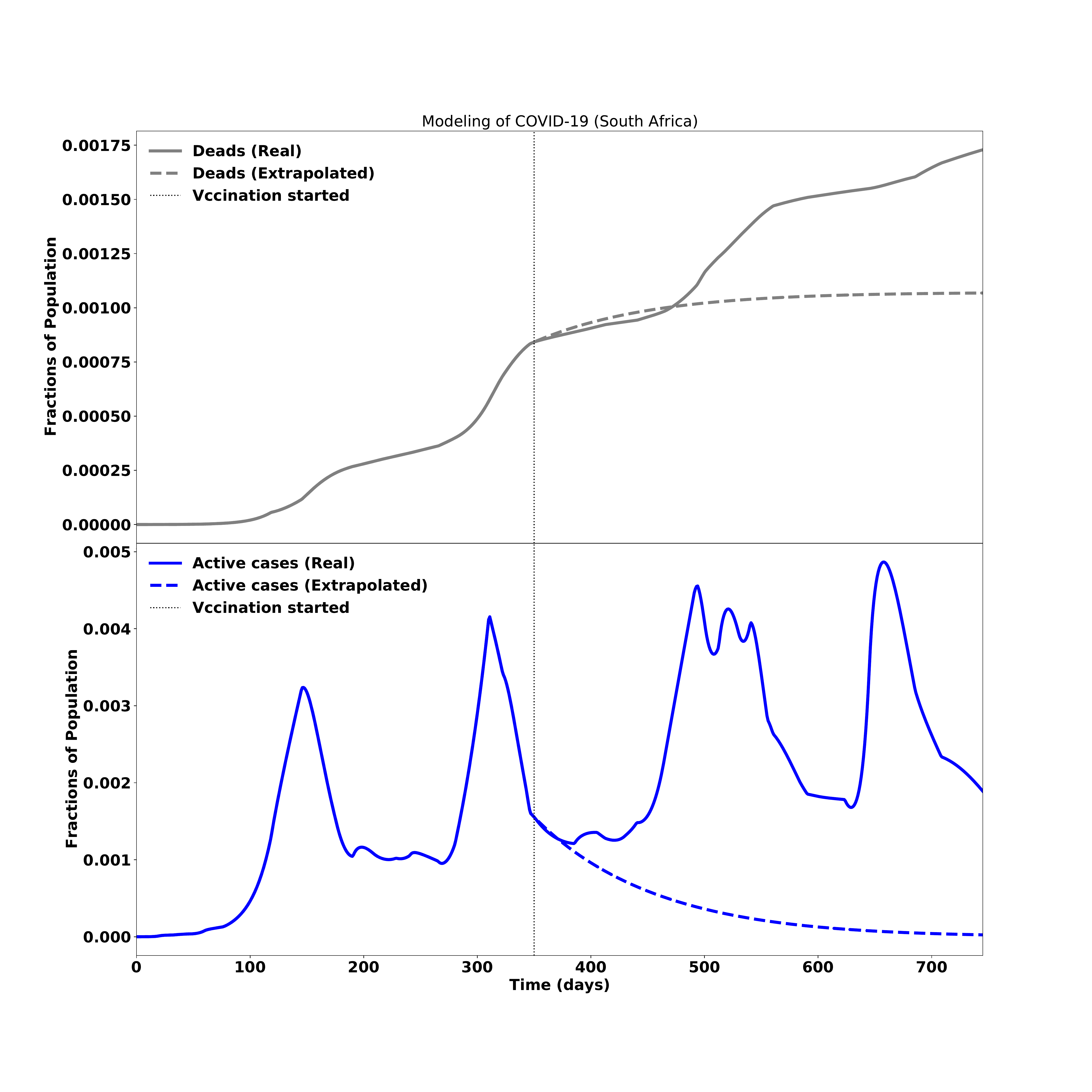}
  	\caption{Death cases (top plot) and active cases (bottom plot) with extrapolation into the period of the vaccination campaign, for South Africa. The plot is shown for a period of 2 years of COVID-19. Day~0 is March 5, 2020. The vertical dotted-line indicates the start of the vaccination campaign.}
   	\label{fig:SA-vac}
  \end{center}
\end{figure}

In this study, we focused on the second year of the COVID-19 pandemic with vaccination rollouts. To discuss the impact of vaccination, we took the case of South Africa where the available data was statistically significant as described in Section~\ref{sec:sa}.  At the beginning of the vaccination campaign, around Day 349 (February 18, 2021), as shown in Figure~\ref{fig:SA-vac} (bottom plot), the number of active cases was declining and the $R_0$, estimated from the bottom-left plot of Figure~\ref{fig:SouthAfricaKenya}, was 0.99 and the government relaxed the control measures to alert level one on March 1, 2021. The SIDARTHE-V model extrapolation into the period of vaccination is shown in Figure~\ref{fig:SA-vac} (dashes curves) and suggests that the active cases should dwindle and the death rate should plateau over time. The relaxation of the control measures without enough vaccinated individuals to reach herd immunity led to the third and fourth waves seen in Figure~\ref{fig:SA-vac} (bottom plot), although vaccination was ramping up (Figure~\ref{fig:SouthAfricaKenya}, top-left plot). 

\begin{figure}[!h]
 \begin{center}
 	\includegraphics[width=\textwidth]{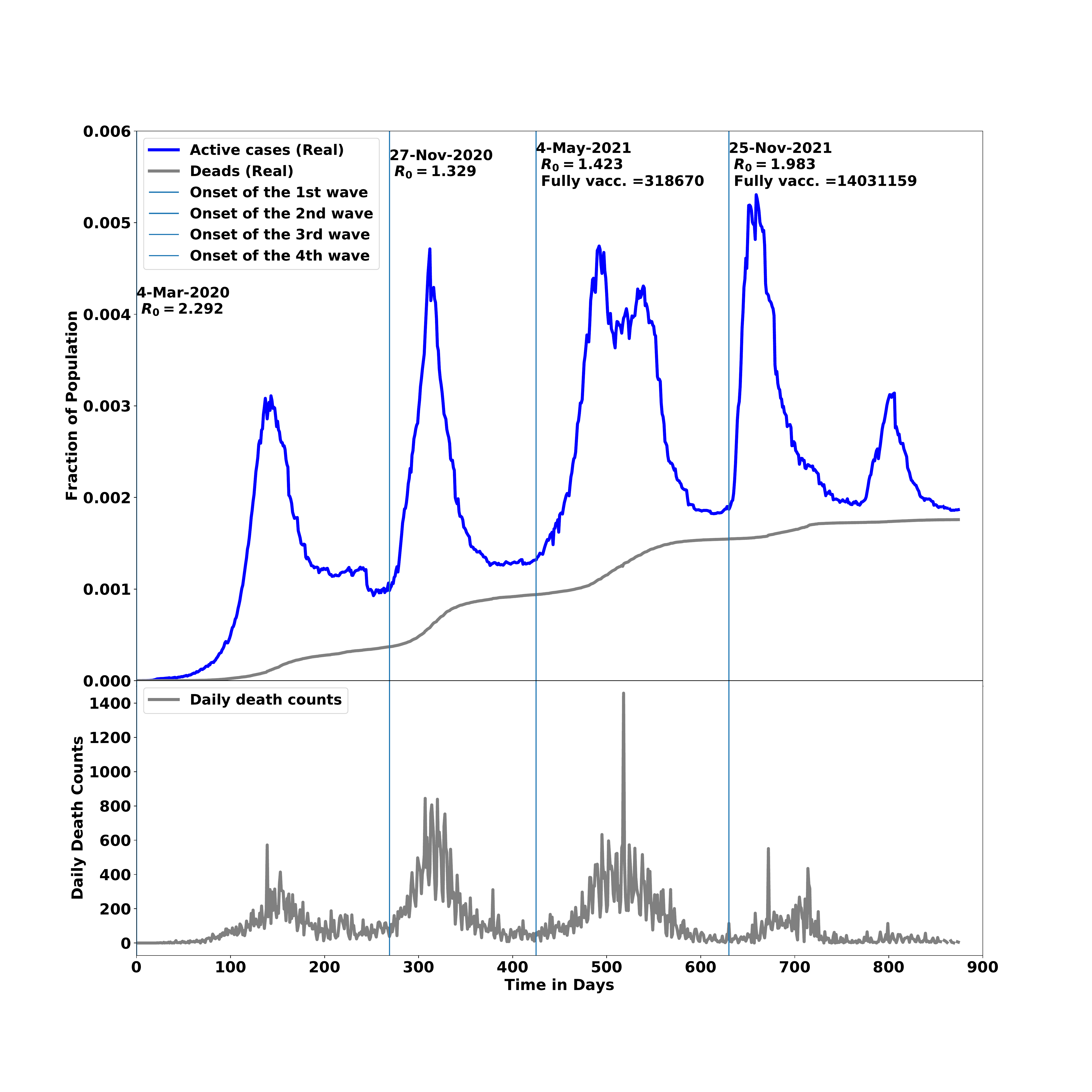}
  	\caption{South African waves of COVID-19 and $R_0$ estimates at the beginning of each wave, numbers of fully-vaccinated persons at the onsets of the third and fourth waves, and the cumulative deaths as a function of time (top plot).  The number of daily death counts is shown in the bottom plot. Day~0 is March 5, 2020. The vertical lines indicate the beginning of pandemic waves.}
   	\label{fig:SA-waves}
  \end{center}
\end{figure}

The number of people $n$ to vaccinate to reach herd immunity is 
\begin{equation}
    n = N \times (1-1/R_0),
    \label{eq:pfrac}
\end{equation}
where $N$ is the population. At the onsets of the third and fourth waves, $R_0$ was estimated at $\sim1.4$ and $\sim2.0$ respectively, as shown in Figure~\ref{fig:SA-waves} (top plot). Assuming $N=60$~million for South Africa, the number of people to vaccinate at the beginning of the third and fourth waves were $n_1=17.1$~million and $n_2=30.0$~million respectively; however, the corresponding numbers of full-vaccinated persons were 318670 and 14031159. Although the vaccination was continued as shown in Figure~\ref{fig:SouthAfricaKenya} (left plot), herd immunity was not reached. 

The lack of herd immunity may be the cause of the fifth wave shown in the top plot of Figure~\ref{fig:SA-waves}. The impact of vaccination was beginning to be felt at the time of the fifth wave---this can be seen in:
\begin{itemize}
\item the fifth wave which was relatively smaller than the previous ones;
\item the cumulative deaths which were plateauing (top plot of Figure~\ref{fig:SA-waves});
\item the daily death counts which had fallen (bottom plot of Figure~\ref{fig:SA-waves});
\item and the relaxation of control measures to level one without resurgence of any significant wave.
\end{itemize}

The impact of vaccination, inferred from Figure~\ref{fig:SA-waves}, appears consistent with what we would expect the vaccination program to achieve. The vaccination program is expected to reduce COVID-19 hospitalizations and deaths. Failure to implement vaccination programs on time can significantly contribute to a rise in the number of infections, increasing the number of hospitalizations and deaths. The basic reproduction number, $R_0$, combines many effects---captured in the model parameters that appear in Equation~\ref{EQ02}---to provide an understanding of the pandemic evolution with control measures or vaccination impacts. These included the death rates ($\tau_1$, $\tau_2$) and worsening rates of infected population $\mu$ and $\nu$. Comparing death rates before and after vaccination, shown in Figure~\ref{fig:SA-waves}, we infer that the parameters $\tau$ are reduced after vaccination campaigns; it means that we can have large infection rates without people dying in large numbers, see the bottom plot of Figure~\ref{fig:SA-waves}. Further, the reduction on the parameters $\mu$ and $\nu$ would reduce severity of infections (see the fifth wave in the top plot of Figure~\ref{fig:SA-waves}) and help in reducing the number of deaths.

The number of death could have been drastically reduced had the non-pharmaceutical interventions been implemented for a while at the beginning of the vaccination program. This is on top of the observation that the death rates due to COVID-19 in Africa are relatively low.

From the data and model simulation, we conclude that vaccination is an important tool in the fight against COVID-19, and early implementation of the vaccination program could have saved a lot of lives.

\section{Conclusions}
\label{sec:conc}
We studied the impact of vaccination in Nigeria, South Africa, Kenya, Ghana, Togo, Mozambique, and Zambia. The SIDARTHE-V model was used in simultaneous fits to active, recovered, extinct, and vaccinated cases in the countries considered. We observed that it is important to combine vaccination roll outs with control measures to contain the pandemic until herd immunity is achieved. To have a better understanding of the effect of the vaccination program in Africa, we studied the South African case in more detail since it was the most impacted country in the continent, and also where we have more statistically significant vaccination data. The impact of vaccination was observed after almost one year when $\sim$ a third of the population had been fully vaccinated. This was reflected in the significantly reduced daily death counts, the plateauing of the cumulative death rate, and the relaxation of control measures without the resurgence of COVID-19 peak waves. For the other countries studied, the impact of vaccination was not easy to gauge because of the relatively smaller numbers of COVID-19 cases and fully vaccinated people. However, the conclusion reached in the South Africa case may be applicable to other countries, that is, vaccination roll outs need to be combined with control measures until enough population has been vaccinated such that the relaxation of control measures no longer lead to significant waves.
 model was developed and analysed to quantify early COVID-19 outbreak transmission in South Africa and explored vaccine efficacy scenarios. It was observed that a vaccine with 70\% efficacy had the capacity to contain the COVID-19 outbreak at a vaccination coverage of 94.44\%; a vaccine with 100\% efficacy required a 66.10\% coverage. Social distancing measures put in place have so far reduced the number of social contacts by 80.31\%. Their results suggested that a highly efficacious vaccine would have been required to contain COVID-19 in South Africa. Therefore, social distancing measures to reduce contact remained key in controlling infections in the absence of vaccines and other therapeutics. 
\newpage

\noindent 

\bibliography{references}

\newpage

\section*{Supplementary Material}

\renewcommand{\thefigure}{SM1}
\begin{sidewaysfigure}
 \begin{center}
 	\includegraphics[scale=0.5]{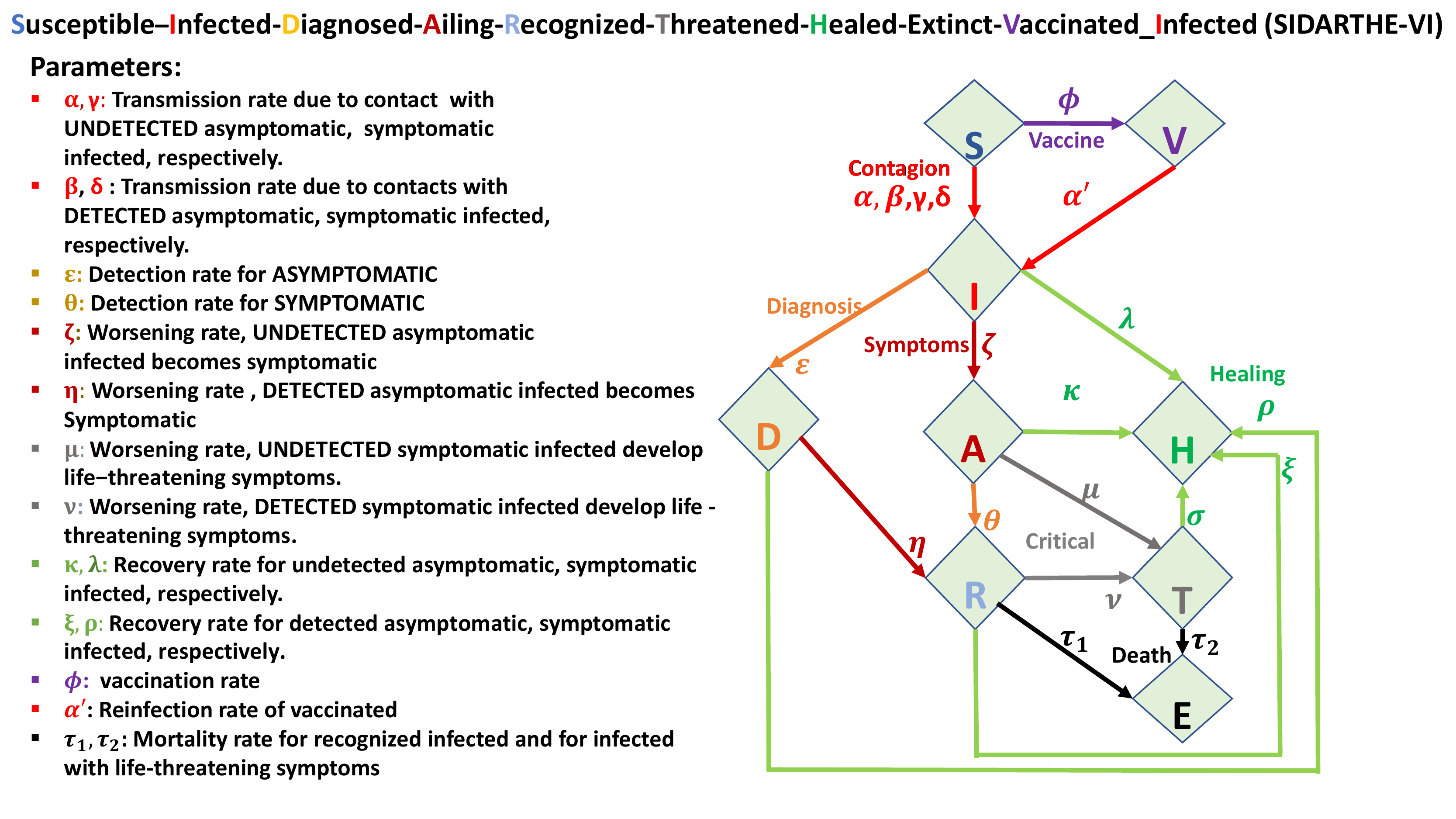}
   	\caption{Flow-chart representing the modified SIDARTHE-V model considering vaccination roll outs; we extended the original SIDARTHE-V of Ref.~\cite{Giordano2021} with the possibility that  vaccinated individuals may become infected.}
   	\label{fig:SIDARTHEV} 
  \end{center}
\end{sidewaysfigure}

\renewcommand{\thefigure}{SM2}
\begin{figure}[!htbp]
 \begin{center}
 	\includegraphics[width=\textwidth]{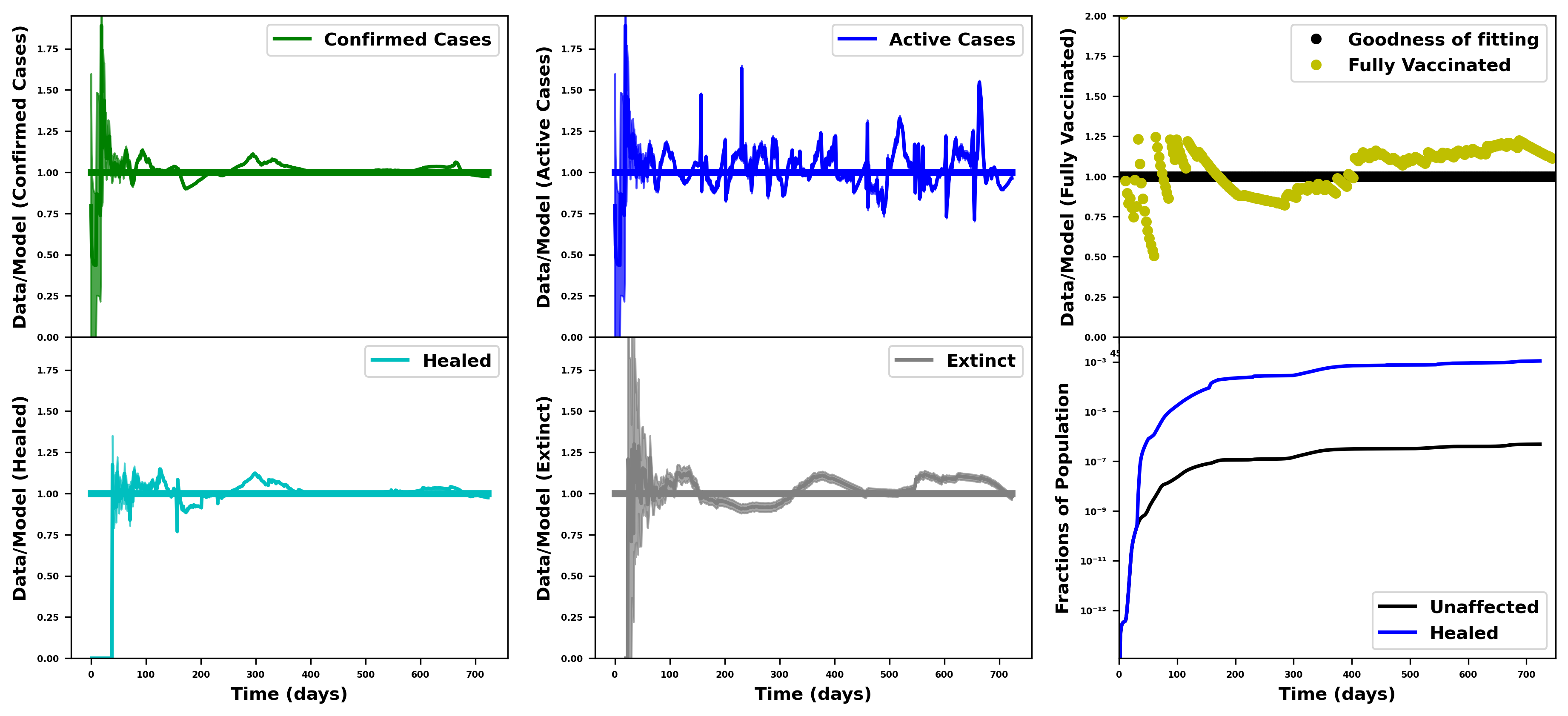}
  	\caption{The goodness-of-fit of the COVID-19 data modelling of Nigeria for confirmed, healed, active, extinct, and fully-vaccinated cases. The bottom-right plot shows a model prediction of the recovered population; also shown in the bottom-right plot, is the non-diagnosed fraction of the people that were infected and recovered without symptoms---this fraction, called the unaffected cases, is not measured or included in the data.}
  	   	\label{fig:Nigeria2}
  \end{center}
\end{figure}

\renewcommand{\thefigure}{SM3}
\begin{figure}[!htbp]
 \begin{center}
 	\includegraphics[width=\textwidth]{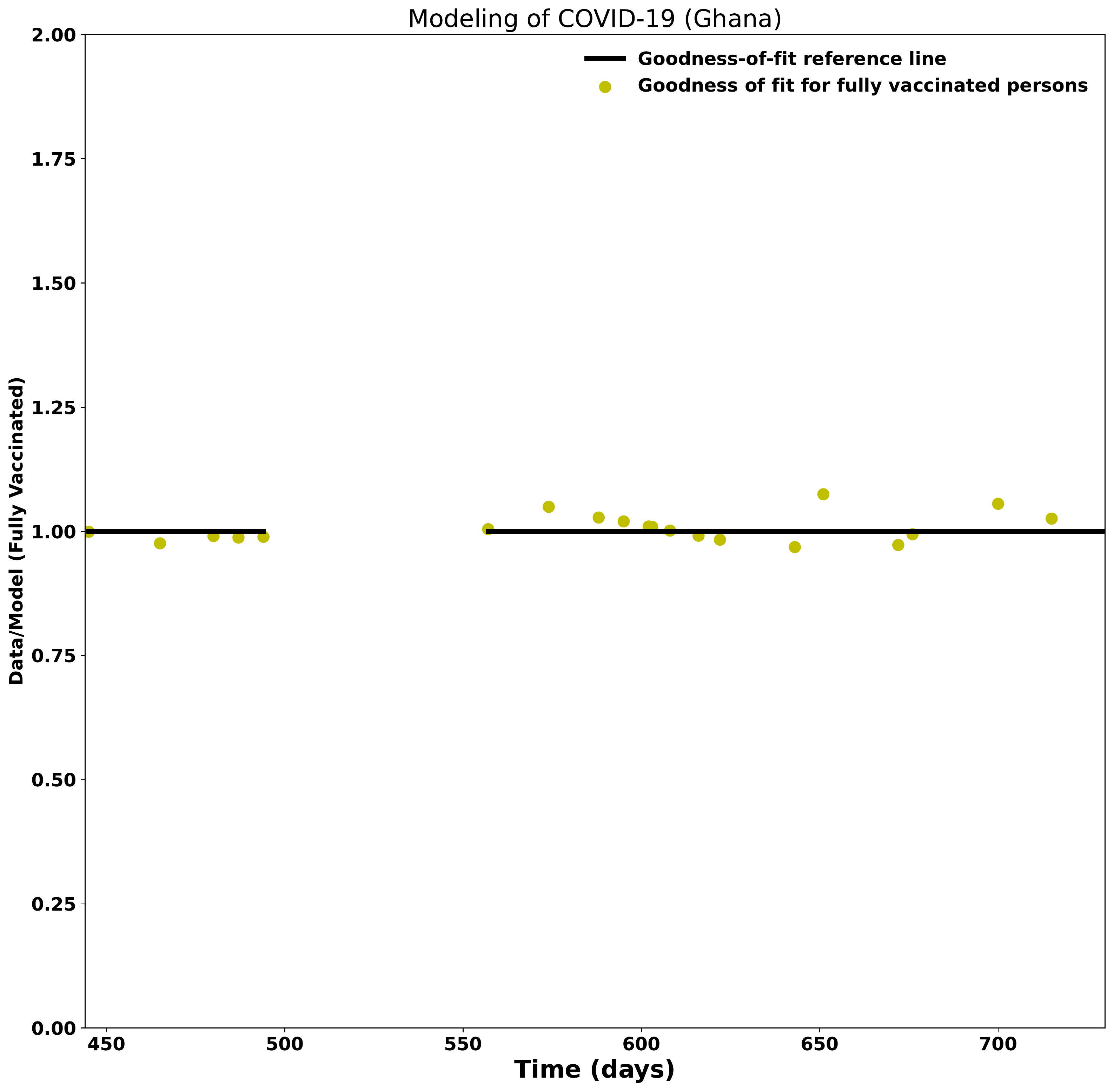}
  	\caption{The plot showing the goodness-of-fit of the COVID-19 data modelling of Ghana for fully-vaccinated individuals over time in days from March 1, 2021, to February 28, 2022.}
   	\label{fig:Ghana2}
  \end{center}
\end{figure}

\renewcommand{\thefigure}{SM4}
\begin{figure}[!htbp]
 \begin{center}
 	\includegraphics[width=\textwidth]{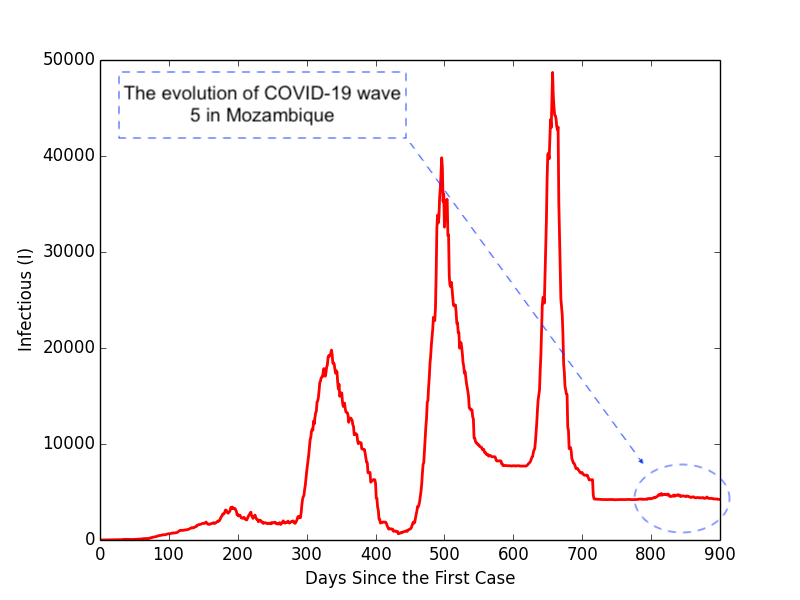}
  	\caption{COVID-19 pandemic waves in Mozambique. Day~0 corresponds to March 20, 2020. The fifth wave occurred during the vaccination campaigns and was relatively smaller than the previous ones.}
 	\label{fig:Mozambique3}
  \end{center}
\end{figure}

\end{document}